\documentclass[12pt]{article}
\usepackage{graphicx}
\newcommand{\beq}{\begin{equation}}
\newcommand{\eeq}{\end{equation}}
\def\bea{\begin{eqnarray}}
\def\eea{\end{eqnarray}}
\def\nn{\nonumber}

\begin{document}

\title{Avoiding the uncertainty from correlation \\between  
        $|\Delta m_{31}^2|$ and CP phase $\delta $ \\in 
        $\nu_\mu \rightarrow \nu_\mu$ long baseline experiments }

\author{Keiichi \textsc{Kimura}$^{1}$\thanks{E-mail: kimukei@eken.phys.nagoya-u.ac.jp}, 
Akira \textsc{Takamura}$^{1,2}$\thanks{E-mail: takamura@eken.phys.nagoya-u.ac.jp} 
          and Tadashi
          \textsc{Yoshikawa}$^{1}$\thanks{E-mail: tadashi@eken.phys.nagoya-u.ac.jp}\\
{\small \it $^1$Department of Physics, Nagoya University, Nagoya, 464-8602,
Japan}\\
{\small \it $^2$Department of Mathematics,
Toyota National College of Technology}\\ 
{\small \it Eisei-cho 2-1, Toyota-shi, 471-8525, Japan}} 
\date{}

\maketitle

\begin{abstract}
We introduce a new index $I_{\Delta m_{31}^2}$ to find where is the
better setup of the baseline length and energy to avoid as well as possible 
the uncertainty from the correlation between $\Delta
m_{31}^2$ and $\cos\delta $ in $\nu_\mu \rightarrow \nu_\mu$ long baseline
experiments. 
\end{abstract}

Detection of the CP effect in lepton sector (MNS matrix\cite{MNS})
is one of the remaining most important subjects in not only 
elementary particle physics but also particle cosmology.
To confirm the existence of the CP phase, many long baseline
experiments\cite{T2K,NOvA,T2KK,NuMI} 
by using $\nu_\mu \rightarrow \nu_e$ oscillation mode are proposing. 
After finding the CP effects in $\nu_\mu \rightarrow \nu_e$, as the
next step, it will be
an important subject to check whether they are consistent with the
standard model(SM). To do so, we need to measure the CP effects(phase) 
independently by using the other oscillation mode.    
Measuring CP phase by $\nu_\mu \rightarrow \nu_\mu$ mode is going to be more 
important to confirm the SM and to investigate the existing possibility 
of new physics. We have to confirm the consistency and the unitarity
in lepton sector\cite{U-check} by comparing the observables extracted from the
different oscillation modes.  
We are investigating the CP effect in $\nu_\mu \rightarrow \nu_\mu$
mode in our work\cite{KTY}. The dependence of the probability on the
CP phase $\delta $ with the maximal 2-3 mixing $\theta_{23} = 45^\circ
$ is written 
as follows\cite{KTY,Kimura:2002hb,Kimura:2006rp}:
\bea
P_{\mu \mu } &=& A_{\mu \mu }\cos\delta + C_{\mu \mu } 
+ D_{\mu \mu}\cos2\delta \nn \\
&\simeq & A_{\mu \mu }\cos\delta + C_{\mu \mu } 
+ O(\sin\theta_{13}\Delta m_{21}^2),
\eea
where $D_{\mu \mu }$ as a coefficient of $\cos2\delta $ is negligible
because the magnitude should be proportional to the quite small parameter 
$\sin\theta_{13}\Delta m_{21}^2$.  $A_{\mu \mu }$ and $C_{\mu \mu }$
are the quantities determined by the parameters except for the CP
phase $\delta $.  
The effect from  $\cos\delta $ depends on the magnitude of $A_{\mu \mu
}$ so that it is an index to know the CP dependence.   
We discussed where is better set of the baseline length $L$ and the neutrino
energy $E$ to extract the CP effect from
$\nu_\mu \rightarrow \nu_\mu $ experiment and pointed out it favors 
$E < 2 {\rm GeV} $, $L >2000{\rm km } $. At once, we showed it seems to be 
difficult to determine the CP phase because there is 
a correlation between $\Delta m_{31}^2$ and $\cos\delta $ in small $L/E$ 
\cite{KTY,DFMR}.     

In this letter, we introduce a new index $I_{\Delta m_{31}^2}$ to 
look for the better region in $(E,L)$ plane and to avoid the uncertainty 
from $\Delta m_{31}^2$-$\cos\delta$ correlation.  Here we are using 
the following input parameters:  
$\Delta m_{21}^2 = 8.1\times10^{-5} {\rm eV}^2, ~\sin^2\theta_{12}=0.31,
~\sin^22\theta_{23}=1 $, and for an unknown parameter $\theta_{13}$,
the upper bound\cite{Apollonio} $\sin^22\theta_{13}=0.16$ is used. 
In the estimation of the probability $P_{\mu \mu}$, we are using 
the exact solution for the neutrino oscillation in matter
\cite{Kimura:2002hb,Yokomakura}.  

The $\Delta m_{31}^2$-$\cos\delta$ correlation is plotted in Fig.1,
where we assume $(\cos\delta, \Delta
m_{31}^{2~true}) = (0,2.5\times 10^{-3} {\rm eV}^2 )$ as the true values 
and the plotted points show where the probability $P_{\mu \mu }$ 
at the fake values $(\cos\delta^\prime ,\Delta m_{31}^2) $ are almost same 
with $P_{\mu \mu }^{true} $ at true value. 
The figure shows the linear relation between the fake parameters
$\Delta m_{31}^2$ and $\cos\delta^\prime $. 
As we discussed in our previous work\cite{KTY}, there is a relation
between the true value and fake one which are producing same
probability approximately within the $L/E \ll 1000 $ as follows:
\bea
(\left|\Delta m_{31}^{2}\right|-\left|\Delta m_{31}^{2~true}\right|)  
    &=& - 4 J_r \Delta m_{21}^2 (\cos\delta^\prime - \cos\delta ) \\
    &=& - 0.0146 \times 10^{-3}~ (\cos\delta^\prime - \cos\delta ),  
\eea 
where $J_r = \frac{1}{8}
\sin2\theta_{12}\sin2\theta_{23}\sin2\theta_{13}
\cos\theta_{13} \simeq 0.045 $ . 
If the relation are satisfied, it means 
one can not determine the magnitude of CP phase 
without uncertainty. Namely, for the error of
$|\Delta m_{31}^2| $, all range of $360^\circ$ is 
satisfied as the solution. Indeed, even if the error is $1\%$ level,
$|\Delta m_{31}^2| = (2.50 \pm 0.02)\times 10^{-3} {\rm eV}^2$, 
we have to consider the uncertainty.
\begin{figure}[htbp]
\begin{center}
\begin{minipage}[c]{0.42\textwidth}
\includegraphics[scale=0.4,angle=-90]{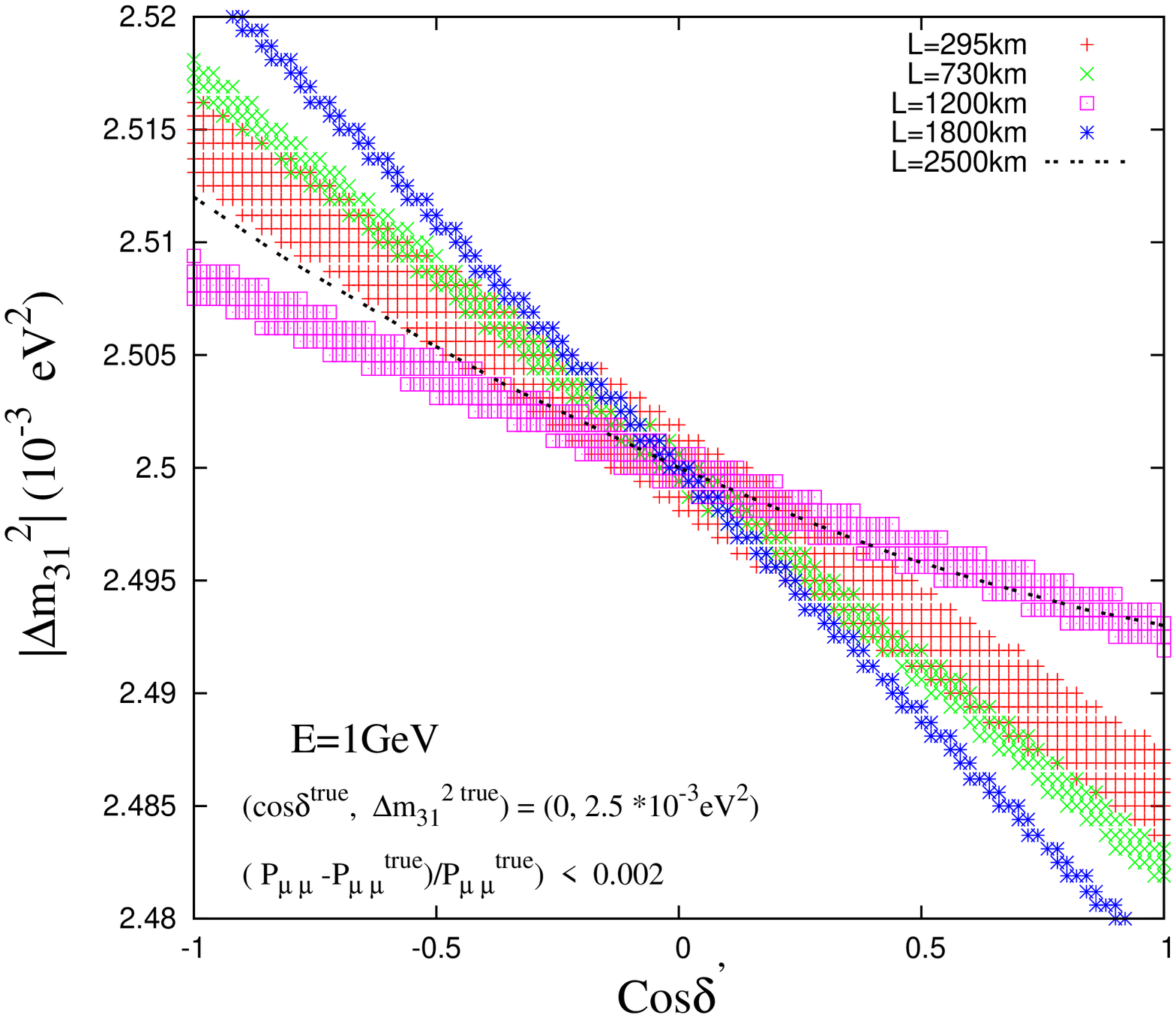}
\end{minipage}
    \hspace*{10mm}
\begin{minipage}[c]{0.42\textwidth}
\includegraphics[scale=0.4,angle=-90]{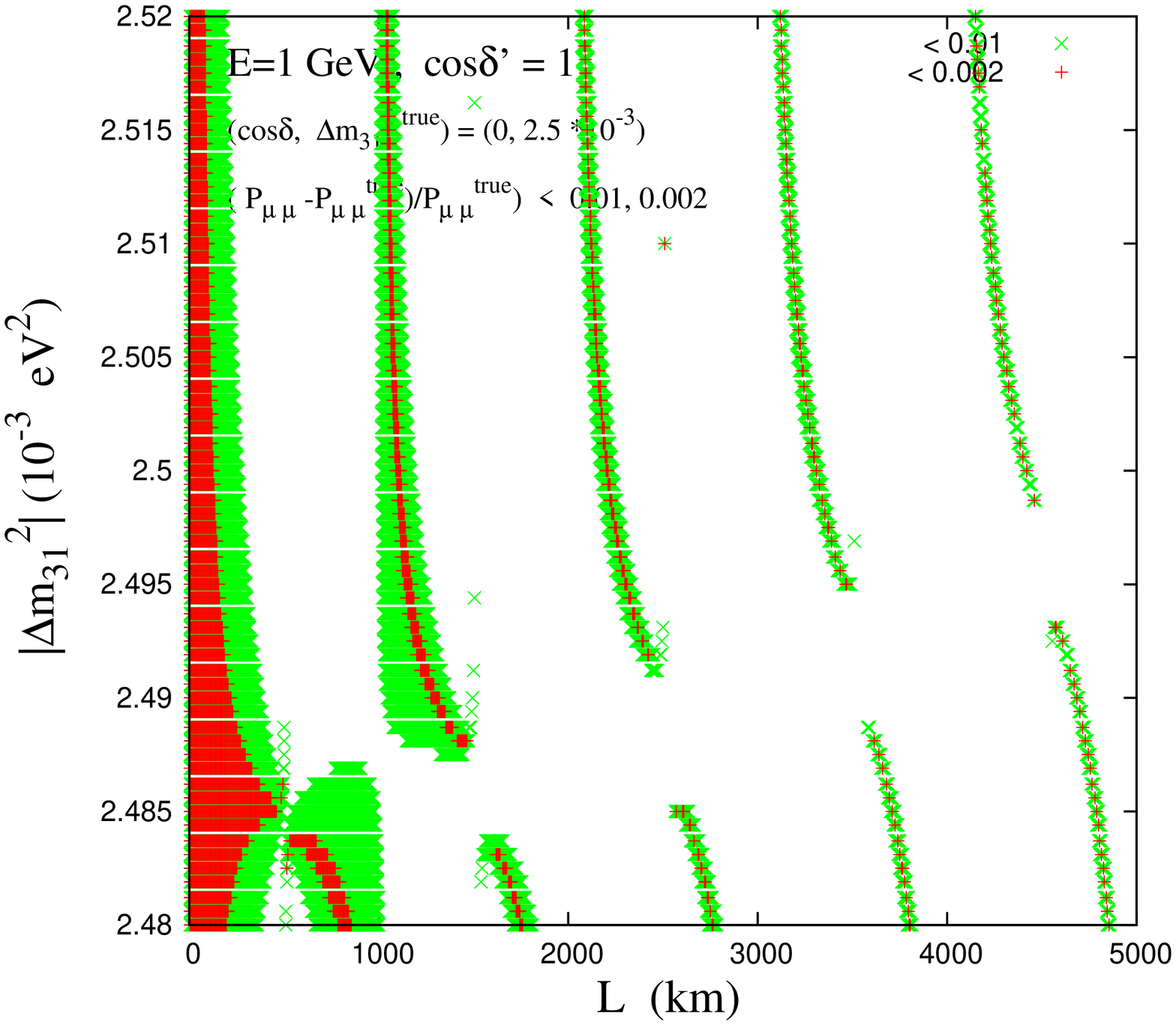}
\end{minipage}  
\caption{ The region shows where   
$(P_{\mu\mu}(\delta^\prime, |\Delta m_{31}^2|) 
  - P_{\mu\mu }^{true})/P_{\mu\mu }^{true} $ is smaller than $0.002$(Left)
and $0.01$,$0.002$(Right) for $\cos\delta^{true}=0$ 
and $\Delta m_{31}^{2~true}=2.5\times 10^{-3}{\rm eV}^2$ at $E=1{\rm GeV}$. 
The left is that for $(\cos\delta^\prime ,\Delta m_{31}^2)$ at several 
baseline length L and the right is for $(L, \Delta m_{31}^2)$ with 
$\cos\delta^\prime = 1. $
}
\end{center}
\end{figure}
{}From the left of Fig.1, one can find the linear relation between 
$|\Delta m_{31}^2|$ and $\cos\delta^\prime $ for several baseline
lengths. The right figure shows the dependence of 
fake region of $|\Delta m_{31}^2|$ on the baseline length $L$ 
at the case of $\cos\delta^\prime
=1 (\delta^\prime = 0^\circ ) $ 
which leads to same probability $P_{\mu \mu }$
with true (input) value $\cos\delta =0 (\delta = 90^\circ
)$. From this, we find that 
the dependence may not be so trivial. 
One can find that the fake region breaks at several $L$s 
in the right of Fig.1.  
Hence we investigate around $L = 505,1000,2000,3000,4000,5000 ({\rm km
})$ where the fake region disappears.            

\begin{figure}[htbp]
\begin{center}
\begin{minipage}[c]{0.42\textwidth}
\includegraphics[scale=0.4,angle=-90]{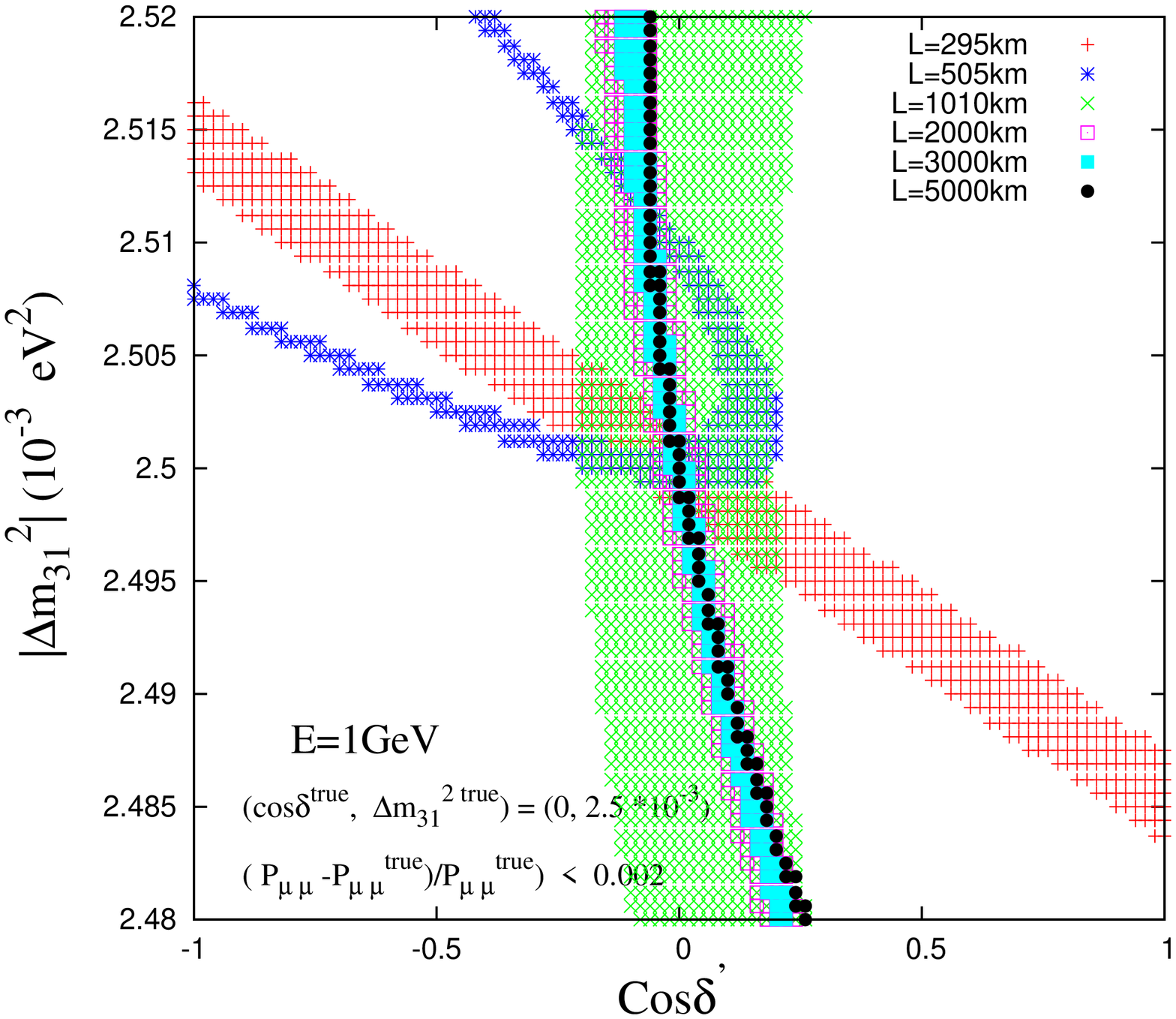}
\end{minipage}
    \hspace*{10mm}
\begin{minipage}[c]{0.42\textwidth}
\includegraphics[scale=0.4,angle=-90]{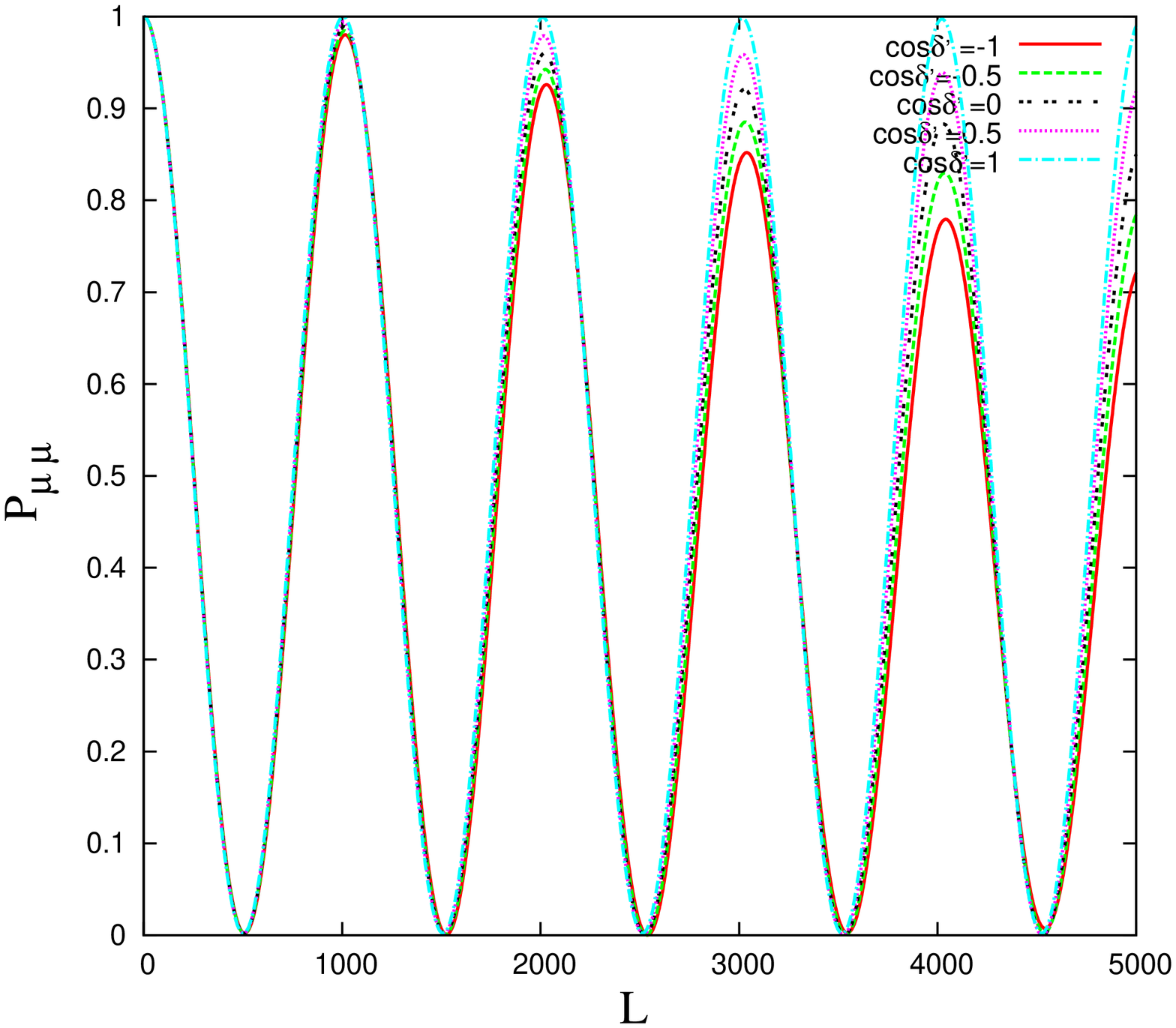}
\end{minipage} 
\caption{ The region satisfing $(P_{\mu\mu}(\delta^\prime, |\Delta m_{31}^2|) 
  - P_{\mu\mu }^{true})/P_{\mu\mu }^{true} < 0.002$ on
$(\cos\delta^{\prime}, \Delta m_{31}^2)$ is shown in the left
figure. The region shows almost same probability with it at
$\cos\delta^{true}=0$ and $\Delta m_{31}^{2~true}=2.5\times 10^{-3}
{\rm eV}^2$ for several $L$s.
The probability as a function of $L$ at
$\delta=0^\circ,90^\circ,180^\circ$ in 
the right.
}
\end{center}
\end{figure}

At $L=~505,1000,2000,3000,4000,5000 ({\rm km })$, 
the correlation of $|\Delta m_{31}^2|$ and $\cos\delta^\prime $ are
plotted in Fig.2(left).  
The dependence seems to be different with the case in Fig.1 
and the almost plotted points are around 
the true value $\cos\delta =0 $ at the $L$ where the fake regions
disappear in Fig.1. It may show that at the several suitable L
one can investigate the CP phase without depending on the error of $|\Delta
m_{31}^2|$ so strong. Where is the region on ($E,L$) ? 
Comparing the right one of Fig.1 with the Fig.2(right) which shows 
the dependence of the probability on the baseline length L, 
where the probability shows maximal or minimal.  
The fake regions also break around $L=500, 1500, 2500, 3500, 4500 ({\rm
km})$ but the probability
is almost $0$ so that we can not extract the CP effect around the $L$. 
On the other hand, at $L=1010,2000,\cdots $, $P_{\mu \mu }$ shows
maximal and the CP effects will also be maximal so that it may be
possible to determine the CP phase without depending on  
$\Delta m_{31}^{2}$ so hard. They correspond to
the region with large $A_{\mu \mu }$. From the left of Fig.2, one can find
the extracted solutions of $\cos\delta^\prime $ is around the true
value for the error of $\Delta m_{31}^{2}$ at the special $L$.       
     
There is the uncertainty in determination of CP phase in 
$\nu_\mu\rightarrow \nu_\mu $ oscillation experiments
because of $\Delta m_{31}^2$-$\cos\delta$ correlation. 
So we introduce a new index $I_{\Delta m_{31}^2}$ 
to search for where is more suitable energy $E$ and distance $L$ 
to avoid the uncertainty. It is defined by the difference of  
maximum and minimum probabilities ($P_{\mu \mu }^{max}$ and 
$P_{\mu \mu }^{min}$)  within the error of  
$\Delta m_{31}^2$ ($|\Delta m_{31}^2| = 2.50\pm0.02 \times
10^{-3} {\rm eV}^2 $)\footnote{We expect that the experimental error
of $\Delta m_{31}^2$ will be reduced up to $1\%$ level 
in the future experiments.}. 
\bea
I_{\Delta m_{31}^2} = \frac{P_{\mu \mu }^{\max}(|\Delta m_{31}^2|)
                           -P_{\mu \mu }^{\min}(|\Delta m_{31}^2|)}
                           {P_{\mu \mu }^{\max}(|\Delta m_{31}^2|)
                            +P_{\mu \mu }^{\min}(|\Delta m_{31}^2|)}, 
\eea
This is the index to indicate how affecting the probability from 
the error of $|\Delta m_{31}^2|$. The regions which the new index is 
as small as possible are favored to avoid the effects from $|\Delta
m_{31}^2|$. On the other hands, to determine the CP phase, the regions 
the dependence on $\cos\delta $ becomes larger are favored. 
Using $A_{\mu \mu }$ one can find the regions. $A_{\mu \mu } $ is a
coefficient of $\cos \delta $ in eq.(1) and it can be also defined as 
the difference between the maximum and minimum of $P_{\mu \mu }$s within all
range of $\delta $.
\bea
A_{\mu \mu } \simeq  \frac{(P_{\mu \mu}|_{\delta=0^\circ} -
                            P_{\mu \mu}|_{\delta=180^\circ})}{2} .
\eea
This corresponds to the numerator of $I_{CP}$\cite{maxcp}.
The region with large $A_{\mu \mu}$ will be useful to extract the CP phase.
The dependence of $I_{\Delta m_{31}^2}$ and $A_{\mu \mu }$ on $L$ at 
$E=1{\rm GeV}$ are plotted in Fig.3.  

\begin{figure}[htbp]
\begin{center}
\begin{minipage}[c]{0.42\textwidth}
\includegraphics[scale=0.4,angle=-90]{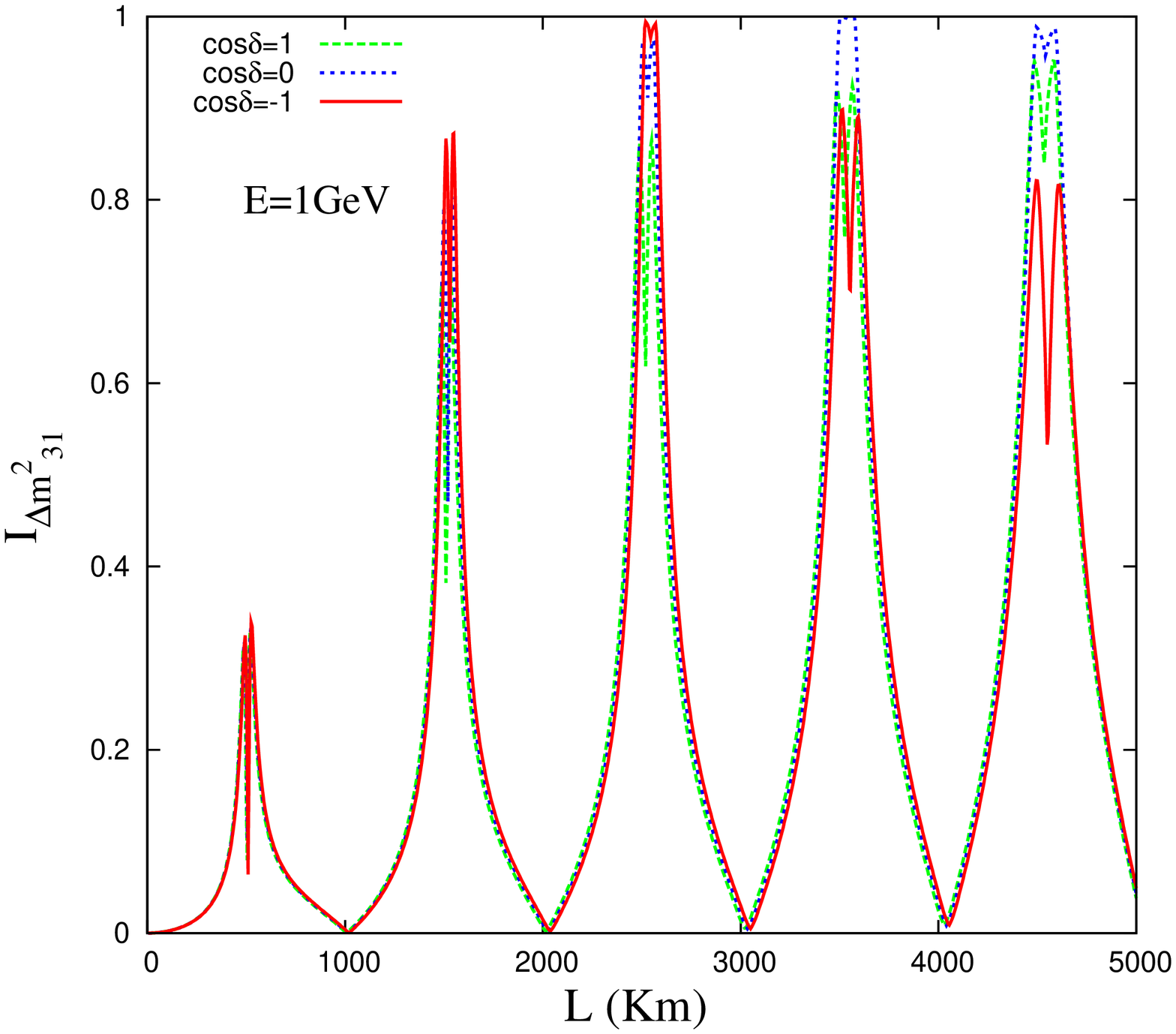}
\end{minipage}
    \hspace*{10mm}
\begin{minipage}[c]{0.42\textwidth}
\includegraphics[scale=0.4,angle=-90]{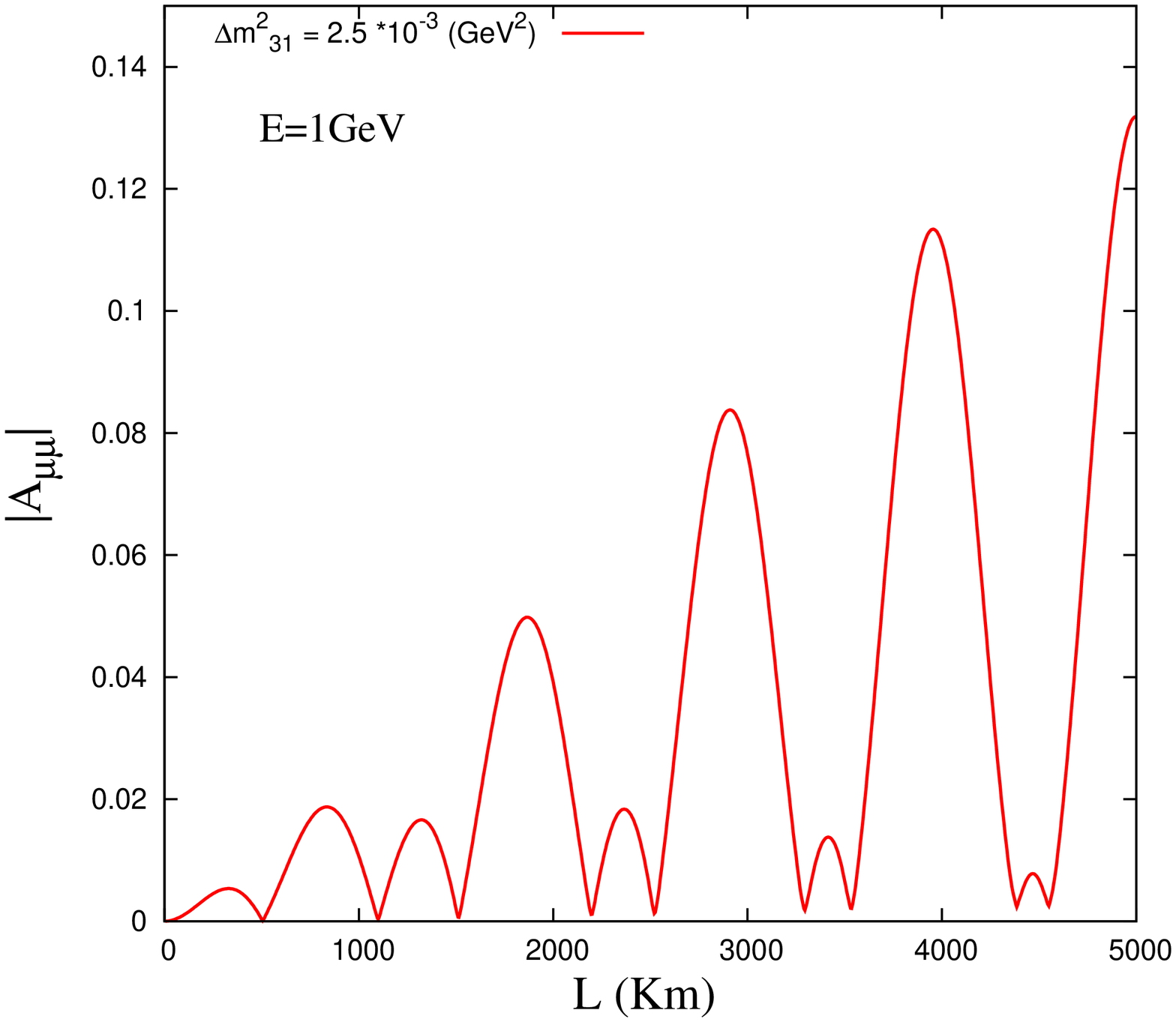}
\end{minipage} 
\caption{ $I_{\Delta m_{31}^2} $ for CP phase $\delta = 0^\circ
,90^\circ ,180^\circ $(left) and $A_{\mu \mu}$(right) 
as the function of baseline length $L$ at $E=1.0 {\rm GeV}$.   
}
\end{center}
\end{figure}

{}From Fig.3, around $1000,2000,3000,4000,5000,...${\rm km}, 
$I_{\Delta m_{31}^2}
$ become minimum and then $A_{\mu \mu}$ are showing nonzero values and
not so small value.  It means that around them, it may be possible to
detect the CP phase without depending on $\Delta m_{31}^2 $ so strong.

{}The same discussion on the $(E,L)$ plane leads to the better 
setup to extract CP angle. In Fig. 4, the region are shown as yellow(red)
area shows $A_{\mu\mu} > 0.01 ~(0.1)$ and gray is $I_{\Delta m_{31}^2} < 0.05 $.
\begin{figure}[htbp]
\begin{center}
\begin{minipage}[c]{\textwidth}
\begin{center}
\includegraphics[scale=0.7]{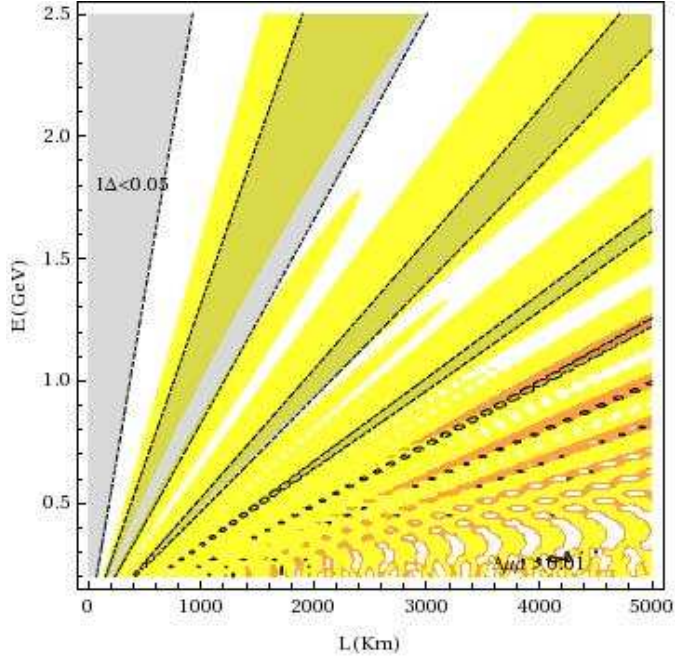}
\end{center}
\end{minipage}
\caption{ The better region to extract CP effect without depending 
on the experimental error of $\Delta m_{31}^2$ are shown 
as the overlapping area. The yellow region show $A_{\mu\mu} > 0.01 $,
the red region is $A_{\mu\mu} > 0.1 $
and the gray one is $I_{\Delta m_{31}^2} < 0.05 $.
}
\end{center}
\end{figure}
{}From Fig.4, one can roughly estimate the better experimental setup
to detect CP phase without depending on the error of $\Delta
m_{31}^2$ so strong.  Around $1000{\rm km}$ which means
T2KK\cite{T2KK}, around $0.5{\rm GeV}$ or 
$1{\rm GeV}$ is better energy region. Indeed, longer $L$ is favored for
$A_{\mu \mu}$ but the minimal values of $I_{\Delta m_{31}^2 }$ will
depart from zero so that we must more carefully  choose the best
place\footnote{We are investigating the T2KK case by using numerical
analysis, which will be reported in the other paper\cite{WP}.}.  
In addition, we define $R_I$ 
as the ratio between $I_{\Delta m_{31}^2}$ and $A_{\mu\mu}$ as following,
\bea
R_I \equiv \frac{I_{\Delta m_{31}^2}}{|A_{\mu \mu }|}.
\eea
Around the $(E,L)$ where the ratio is smaller than 1    
the dependence of $P_{\mu \mu }$ on $\Delta m_{31}^2 $ should be smaller
than the effect by CP phase. In Fig.5, the region of small $R_I$ are
plotted. It may show that we can constrain $\delta $ by using 
the setup of ($E.L$).      

\begin{figure}[htbp]
\begin{center}
\begin{minipage}[c]{0.42\textwidth}
\includegraphics[scale=0.55]{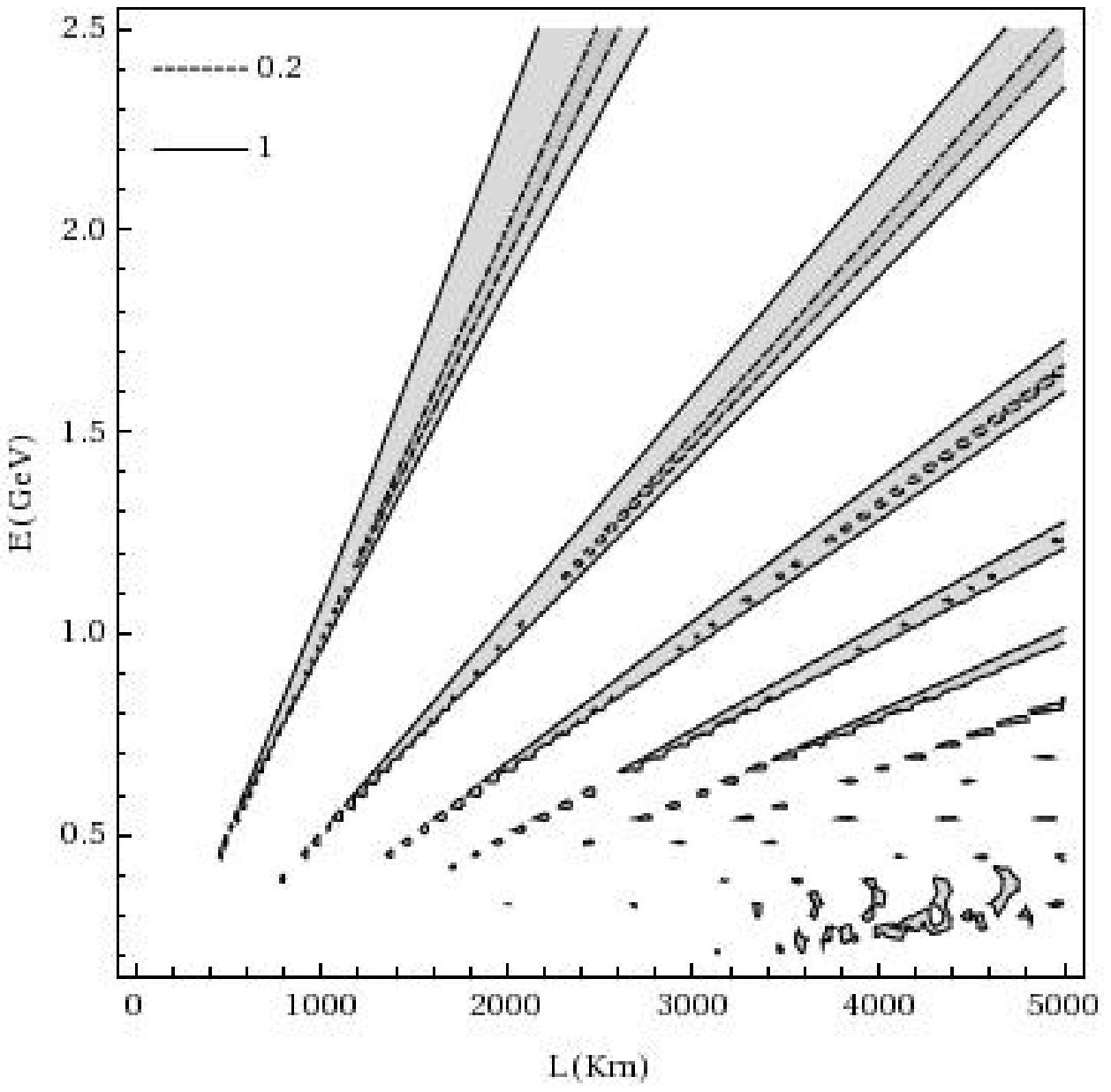}
\end{minipage}
    \hspace*{10mm}
\begin{minipage}[c]{0.45\textwidth}
\includegraphics[scale=0.55]{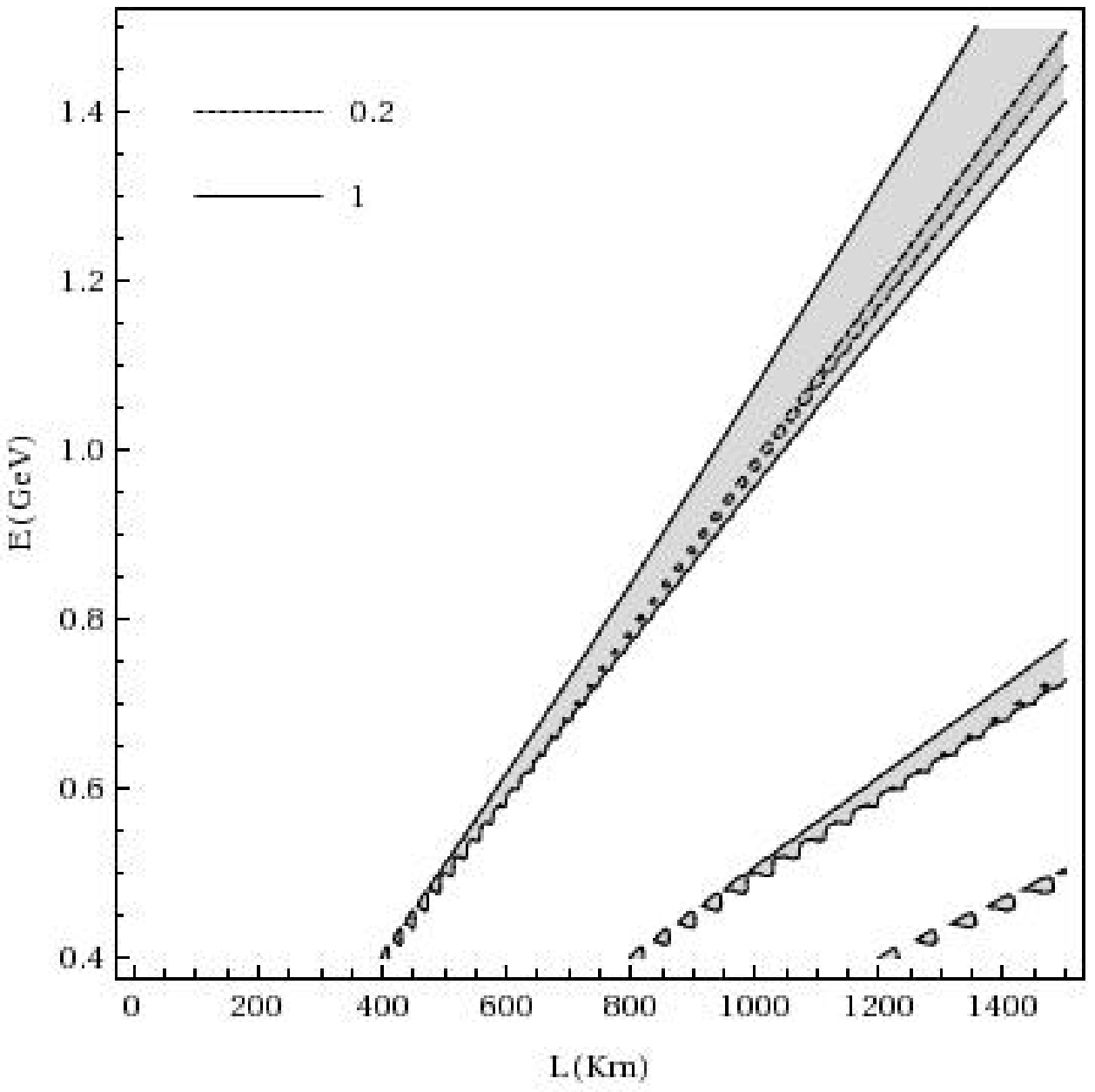}
\end{minipage}
\caption{ The better region to extract CP effect without depending 
on the experimental error of $\Delta m_{31}^2$ are shown 
as the area with $R_I < 0.2 $(dashed lines), and $1$(solid line).
}
\end{center}
\end{figure}

If the experiments are fixed, taking the small and suitable energy 
bin size, we can avoid the uncertainty from $|\Delta
m_{31}^2|$-$\cos\delta $ correlation. 
To estimate where is the better setup of $(E,L)$, 
the new index may be a powerful tool. 
As you find from Fig.1, even if the error of $\Delta m_{31}^2$ is 
reduced, the uncertainty of $\delta $ will remain in almost cases 
which are not chosen as so good $(E,L)$. We expect that the new index
is going to be such powerful tool to improve the determination of CP
phase in $\nu_{\mu } \rightarrow \nu_{\mu }$ oscillation and it will
be possible to confirm 
the consistency with the CP effects in $\nu_{\mu } \rightarrow \nu_{e
}$.      

\section*{Acknowledgment}
The work of T.Y. was supported by 21st Century COE Program of Nagoya 
University.

\end{document}